\documentclass{IOS-Book-Article}

\usepackage{cite}
\usepackage{amsmath,amssymb,amsfonts}
\usepackage{algorithmic}
\usepackage{graphicx}
\usepackage{textcomp}
\usepackage{xcolor}
\def\BibTeX{{\rm B\kern-.05em{\sc i\kern-.025em b}\kern-.08em
    T\kern-.1667em\lower.7ex\hbox{E}\kern-.125emX}}

\usepackage{tabularx}
\usepackage{booktabs}
\usepackage{makecell}
\usepackage{enumitem}
\usepackage{hyperref}
\usepackage{float}
\usepackage{longtable}
\usepackage{booktabs}
\usepackage{fancyvrb}
\usepackage{subcaption}
\usepackage{tabularx}
\usepackage{enumitem}
\usepackage{lineno}

\newcommand{\mhub}{M-Hub}
\newcommand{\INSPIRE}{INSPIRE} 
\newcommand{\INSPIREfull}{INtelligent Scientific Publication Information Retrieval Engine} 

\begin{document}



\begin{frontmatter}

\title{Large Language Model based air quality monitoring and localized alert generation}

\runningtitle{Large Language Model based air quality monitoring and localized alert generation}


\author[A]{\fnms{Ricardo} \snm{Vieira}\thanks{Corresponding author. Ricardo Bastos Leta Vieira E-mail: rvieira@inf.puc-rio.br}},
\author[A]{\fnms{Luis} \snm{Tavares}},
\author[A]{\fnms{Kaylane} \snm{Lima}},
\author[A]{\fnms{Lucas} \snm{de Souza}},
\author[A]{\fnms{Arthur} \snm{Poggy}},
\author[A]{\fnms{João} \snm{Lima}},
\author[A]{\fnms{Vitor} \snm{Pinheiro}}
and
\author[A]{\fnms{Markus} \snm{Endler}}

\runningauthor{R. Vieira et al.}

\address[A]{Departamento de Informática, Pontifical Catholic University of Rio de Janeiro, Rio de Janeiro, Brazil}

\begin{abstract}
Poor indoor air quality can cause up to five times more direct health problems to occupants than outdoor air. In particular, it may cause headaches, fatigue, eye/throat irritation, and long-time exposure is linked to respiratory and heart as well as some forms of cancer. Despite the importance of indoor health and well-being, most current monitoring devices and systems (usually for offices and workspaces) are passive. The Environmental Quality Monitor (EnQyMo) platform is a generic Internet of Things (IoT) middleware designed to process several sensor data related to air quality in indoor spaces and correlate this data with health exposure risks of users/workplace employees. Using Bluetooth Low Energy (BLE) beacons and a mobile IoT middleware it is able to identify the (smartphone) users exposed to these polluted air or high CO2 (carbon dioxide) levels, and generate location-specific alarms only to the users at the places with the unhealthy air conditions. At the core of EnQyMo is an agency of Large Language Models (LLMs) capable of interpreting regulatory standards and scientific literature to automatically identify critical health exposure levels.
\end{abstract}



\end{frontmatter}

\section{Introduction}
\label{sec:intro}
As large fraction of the urban population spends most of its time indoors, the air quality in these spaces critical determinant of the citizen's health. 
Poor Indoor Air Quality (IAQ) may be due to fuel-burning combustion appliances, smoking, building materials and furnishings, such as:
deteriorated asbestos insulation, newly installed flooring or carpets, central HVAC systems,  humidification devices or excess moisture.\footnote{United States Environmental Protection Agency, www.epa.gov/indoor-air-quality-iaq/introduction-indoor-air-quality\#sources} 
On the other hand,  poor IAQ can cause immediate effects like headaches, fatigue, and eye/throat irritation. Long-term exposure is linked to respiratory diseases, heart disease, and, potentially, some forms of cancer. Very often it also causes the exacerbation of chronic respiratory diseases, such as asthma and COPD.  
In spite of importance for health and well-being, most current monitoring solutions remain passive. They provide raw data (e.g., parts per million of $CO_2$ or $\mu g/m^3$ of particulate matter) without translating these values into actionable health insights for the specific individuals at risk. Currently, there is a significant gap in integrated systems that can simultaneously monitor localized indoor environments and provide real-time, context-aware health interventions \cite{Azmeel-2025}.

To address this, we propose EnQyMo (Environmental Quality Monitoring), an intelligent system designed to bridge the gap between environmental sensing and personalized health protection. Unlike traditional rule-based systems that rely on static, universal thresholds, EnQyMo utilizes an agentic Large Language Model (LLM) framework for its ability to parse and "reason" over complex, unstructured scientific literature and medical guidelines. By autonomously retrieving and processing technical documents, the system defines dynamic hazard intervals that correlate specific pollutant levels with their impact on chronic respiratory conditions, effectively acting as a digital health expert.

The EnQyMo architecture leverages the ContextNet \cite{endler:VLIOT:18} and Mobile Hub (\mhub) \cite{Talavera:2015} middleware to solve the challenge of localized alerting. By identifying the precise group of people within a hazardous zone (e.g., a specific office or machine room) and tracking their exposure duration via BLE beacons used for presence tracking, the system delivers targeted notifications. Furthermore, to address the common IoT challenge of data volatility, the system includes a pre-processing layer that filters sensor noise, ensuring that the LLM-driven inference engine receives reliable data for decision-making.

While the long-term vision for EnQyMo includes autonomous environmental control, such as the automated triggering of HVAC systems or filtration units, this paper focuses on the foundational pipeline of sensing, intelligent inference, and targeted notification. We also briefly discuss the privacy-by-design considerations inherent in our middleware choice, ensuring that location data is handled securely to protect user confidentiality.

The remainder of this paper is organized as follows: Section II discusses related work in IAQ and well-being. Section III reviews the underlying ContextNet and Mobile Hub technologies. Section IV details the EnQyMo hardware and software architecture. Section V explains the INSPIRE engine and its LLM-based agentic workflow, including the specific scientific sources utilized. Section VI presents early experimental results, and Section VII concludes the paper.

\section{Related Work}
\label{sec:related}
\begin{table}[t]
    \centering
    \tiny
    \setlength{\tabcolsep}{3pt}
    \renewcommand{\arraystretch}{2}
    \caption{Synthesis of related works and their connection with EnQyMo.}
    \label{tab:survey-enqymo-abnt}
    
    \begin{tabular}{p{0.12\linewidth} *{6}{>{\centering\arraybackslash}p{0.12\linewidth}}}
        \toprule
        \textbf{Aspect} &
        \textbf{Kanagaratnam {\em et al}
\cite{He2025LLMIndoorIAQ}} &
        \textbf{Renold {\em et al}
        \cite{Renold2025IoTClassroomAQ}} &
        \textbf{Arslan {\em et al}
        \cite{Arslan2025AgenticRAGLLM}} &
        \textbf{Pan {\em et al}
        \cite{PanNipu2025MCP}} &
        \textbf{Li {\em et al} \cite{LinHua2025PILLM}} &
        \textbf{EnQyMo (Proposed)} \\
        \midrule
        
        \textbf{Main Focus} &
       Robotic mobile IAQ monitoring &
        Predictive IAQ modeling &
        Monitoring indoor environmental conditions &
        LLM-enhanced IAQ with contextual data integration &
        HVAC fault detection \& control &
        LLM-based IAQ monitoring and localized alert generation \\
        
        \textbf{LLM / AI} &
        GPT &
        Deep Learning (LSTM) &
        \shortstack{LLaMA and\\Agentic RAG} &
        DeepSeek &
        \shortstack{Physics-based\\LLM} &
        \shortstack{Specialized LLM\\Agents (RAG)} \\

        \textbf{IoT Application} &
            \shortstack{Mobile robot\\(PM/CO\textsubscript{2})} &
            \shortstack{Classroom\\sensor nodes\\(IAQ+dust)} &
            \shortstack{Building sens.\\+ BIM \\ integration} &
            \shortstack{IoT sensors\\(T/ RH/\\ CO\textsubscript{2}/PM)\\via MCP/API} &
            \shortstack{HVAC/BMS\\telemetry\\(fault data)} &
            \shortstack{IAQ nodes\\+ BLE beacons\\+ phone hub} \\

        \textbf{Hub Device} &
       Mobile Robot & Fixed Sensor Node & Software Platform & Web Service & Analysis Framework & Mobile Smartphone \\

        \textbf{Implementation Maturity Level} &
        \shortstack{Partial} &
        \shortstack{Complete} &
        \shortstack{Complete} &
        \shortstack{Complete} &
        \shortstack{Complete} &
        \shortstack{Partial} \\

        \textbf{Localized Alerts} &
        No & Yes & No & Yes & No & Yes \\
        
        \textbf{Real-Time Alerts} &
        No & No & No & Yes & No & Yes \\
        \bottomrule
    \end{tabular}
\end{table}

Current studies reinforce the importance of a comprehensive vision that unites thermal comfort, air quality, and occupational health, highlighting how the modularity and scalability of monitoring systems are essential for their adaptation to different built contexts \cite{ElLeathey-2023}. This ecosystem of research, each addressing specific challenges with specialized tools, constitutes a rich and well-established panorama that serves as an essential starting point for the development of even more comprehensive and results-oriented solutions, such as the EnQyMo platform proposed in this work.

The application of Artificial Intelligence to Indoor Air Quality (IAQ) encompasses multiple stages of the analytical value chain. A robotic inspection system \cite{He2025LLMIndoorIAQ}, for instance, employs Large Language Models (LLMs) to optimize data acquisition, automating the spatial mapping of pollutants. This approach primarily focuses on the automated generation of high-resolution environmental diagnostics. EnQyMo, in contrast, positions its contribution in the subsequent stage of semantic interpretation and contextual action. While the robotic system focuses on solving detection and localization problems by establishing the identity and spatial distribution of contaminants, the EnQyMo architecture advances toward health risk inference. Its machine learning core employs LLM agents to correlate multi-spectral pollutant data, collected through a distributed sensor network, with structured biomedical knowledge. This capability enables the system to transcend mere environmental monitoring, translating real-time measurements into personalized health alerts and targeted interventions for effectively exposed individuals, representing a conceptual evolution from diagnostic systems to proactive health protection systems.

This health-oriented specialization conceptually distinguishes EnQyMo from other AI-based platforms, such as ThermalComfortBot \cite{Arslan2025AgenticRAGLLM}, which implements a Retrieval-Augmented Generation (RAG) architecture to integrate Building Information Models (BIM), sensor streams, and LLMs with a focus on optimizing thermal comfort. EnQyMo, however, significantly broadens the scope of AI applied to IAQ by transcending the domain of environmental comfort. Its machine learning core is specifically designed to identify correlations between complex multimodal pollutant profiles and respiratory health conditions, such as allergies or chronic diseases. Thus, the system’s LLM agents operate as an advanced environmental diagnostic mechanism, generating not only descriptive reports but also proactive occupational alerts grounded in individual and collective exposure profiles.

Within the scope of AI-enhanced IAQ solutions, conversational interfaces have also emerged as a promising direction to make environmental data more accessible to non-expert users. Pan and Nipu \cite{PanNipu2025MCP} (2025) propose an LLM-enhanced air quality monitoring interface that uses the Model Context Protocol (MCP) to integrate real-time IoT sensor data. The system enables natural-language queries and produces contextualized, accurate responses, significantly reducing model hallucinations \cite{PanNipu2025MCP}. While this approach advances accessibility and interpretability, its primary emphasis remains on the interface layer and on communicating information. EnQyMo extends this concept by employing LLM agents not only for interpretation, but as the core of a proactive health-risk inference system. Whereas Pan and Nipu’s solution optimizes human--machine interaction for environmental queries, EnQyMo leverages a similar AI paradigm (tool-using LLM agents via standardized protocols) toward a distinct objective: correlating multimodal pollutant signals with biomedical knowledge and issuing personalized health alerts and contextualized interventions. This represents an evolution from diagnostic systems to proactive, risk-oriented health protection systems.

The pursuit of explainable and physically consistent AI models represents another front of advancement in the field, exemplified by the Physics-Informed Large Language Model (PILLM) framework by Subin Lin and Chuanbo Hua (2025), which embeds HVAC thermodynamic and control-theoretic constraints into LLM-driven rule generation for transparent fault detection \cite{LinHua2025PILLM}. EnQyMo aligns with this principle of prioritizing explainability and domain knowledge integration. However, its ML implementation emphasizes a different end goal: coupling contextualized reasoning (via its hub and orchestration modules) to an occupant-aware monitoring paradigm. In this setting, explainability supports not only diagnosing IAQ degradation and its likely physical drivers, but also justifying selective notification actions—i.e., clarifying why a given alert was issued for a specific person, in a specific zone, at a specific time.

Finally, in contrast to purely data-driven models that employ deep learning techniques such as hierarchical clustering and LSTM networks to uncover pollutant patterns and predict their concentrations \cite{Renold2025IoTClassroomAQ}, EnQyMo’s AI architecture introduces an additional layer of semantic reasoning. The system complements the detection of purely statistical patterns with an agentive capacity to interpret such patterns in light of established scientific knowledge. Consequently, the platform not only predicts pollutant concentration trajectories but also proactively assesses the health implications of these conditions. Through its low-cost IoT infrastructure, it autonomously operates to mitigate risks in a targeted and contextually relevant manner.

The analysis presented in Table~\ref{tab:survey-enqymo-abnt} shows that EnQyMo, although still under development, offers a uniquely integrative perspective within the intelligent IAQ monitoring landscape. While existing solutions demonstrate maturity in specific domains such as predictive modeling, conversational interfaces, or anomaly diagnostics, our architectural proposal advances by coherently combining capabilities that remain largely disconnected in the current literature. EnQyMo’s distinction lies in integrating a smartphone-based mobile hub, specialized LLM agents for continuous interpretation of scientific knowledge, and a closed-loop mechanism for generating personalized, actionable alerts. Its goal extends beyond environmental characterization toward proactive, risk-based health protection centered on the individual.

The comparison further indicates a natural trade-off between scope and maturity. Systems with a narrower focus tend to achieve higher operational maturity, whereas broader and more integrated proposals, such as EnQyMo, are at earlier stages of realization.

\section{Previous Works and Underlying Technologies}
\label{sec:underlying}

\subsection{ContextNet Kafka Core}
The ContextNet Kafka Core (CKC)\cite{endler:VLIOT:18} comprises four  backend middleware microservices, that together support scalable  IoT communication, sensing and actuation through mobile edge devices (that we named \mhub s). 
These CKC services execute on servers or virtual machines in a cloud, and are: (i)  the Gateway, that efficiently handles up to thousands of connections with \mhub s, (ii) the GroupDefiner, which dynamically creates groups of \mhub s with some context (e.g. their current location), (iii) the PoA Manager, which tries to balance the load of the other CKC services, and (iv) the  Mobile Temporary Disconnect (MTD), which is informed of temporary disconnections of some Mobile Hub withholds edge-bound messages until the Mobile Hub device reconnects with the system.

\begin{figure}[H]
    \includegraphics[width=1.05\linewidth]{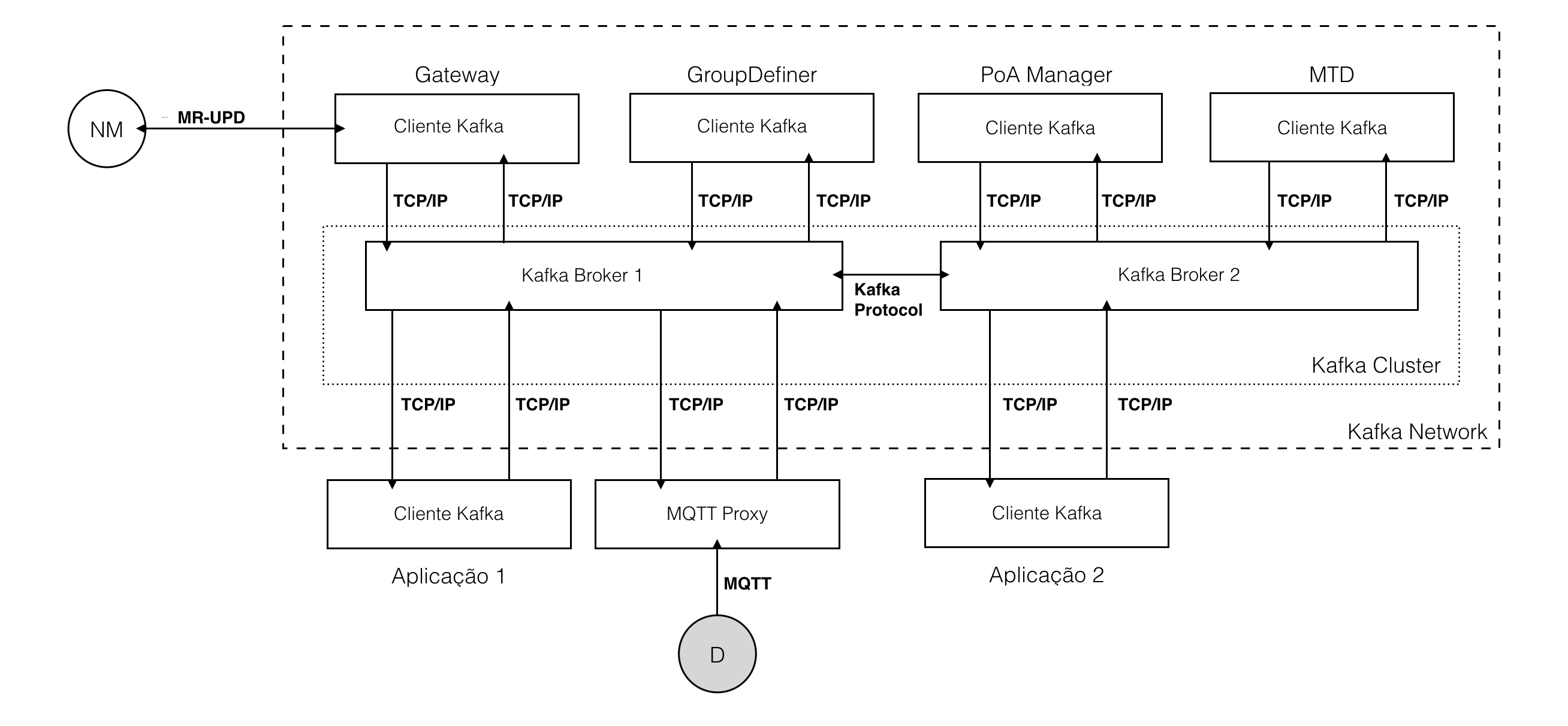}
    \caption{Diagram of CKC architecture. mobile nodes are represented by the element “MN”}
    \label{fig:ckc_diagram}
\end{figure}

In figure \ref{fig:ckc_diagram}, mobile nodes connect to CKC through the Gateway via an MR-UDP connection. The Gateway, which is a Kafka Client, communicates via TCP/IP with one of the Kafka servers (Brokers). The Kafka Brokers make up the Kafka Cluster. The Cluster communicates via TCP/IP with the other CKC microservices: PoA Manager, Group Definer, and Mobile Temporary Disconnect, which are Kafka Clients. The applications (“Application 1” and “Application 2”) developed on the middleware are also Kafka Clients and also communicate with the Cluster via TCP/IP. \cite{endler:VLIOT:18}

\subsection{Mobile Hub}
The Mobile Hub (\mhub)\cite{Talavera:2015} is a general-purpose middleware service (executed on conventional smartphones) that discovers, registers, and enables remote unicast and group-cast communication with several kinds of peripheral beacons or IoT devices, connecting them to the CKC. Hence, the \mhub s thus "bridges the gap" between the Internet connection CKC, and the short-range WPAN  connections established with the peripheral IoT devices, which may be very simple wearable devices, sensor devices, smart bulbs, door locks,   robots with embedded sensors or actuators, none of them with major processing and storage capacity. 

\subsection{N8N}

The physical implementation of the AI Platform for Health utilizes N8N, a workflow automation tool, to orchestrate the data processing pipeline. The implementation is structured into three distinct stages:

\begin{enumerate}
    \item \textbf{Data Historization:} The first step focuses on recording all data received from the sensors into a historical database. This ensures a persistent record of environmental conditions for future training and auditing.
    
    \item \textbf{Knowledge Base Construction (RAG):} This step involves the vectorization of scientific documents to support the AI agents. By ingesting scientific papers and technical standards, the system builds a vector store that allows the agents to perform Retrieval-Augmented Generation (RAG), facilitating accurate correlations between specific pollutants and health risks.
    
    \item \textbf{Monitoring and Alerting Loop:} The system executes a periodic check (configured at specific intervals) of the sensor measurements. It queries the historical metrics and triggers the AI agents for inference. If an anomaly or risk factor is identified, the workflow generates a structured message to be dispatched to ContextNet, ensuring real-time notification.
\end{enumerate}


    


\section{Platform for Air Quality Monitoring and Alert Generation: an Overview}
\label{sec:system}
The EnQyMo system was developed in cooperation with 3R Brasil, a local company that develops and commercializes high-fidelity sensors for measuring environmental pollution of any kind. This company was the provider of the Air Quality Sensor 
Device.

\subsection{Air Quality Sensor 
Device}
\begin{figure}[htbp]
    \centering
    \includegraphics[width=0.55\linewidth]{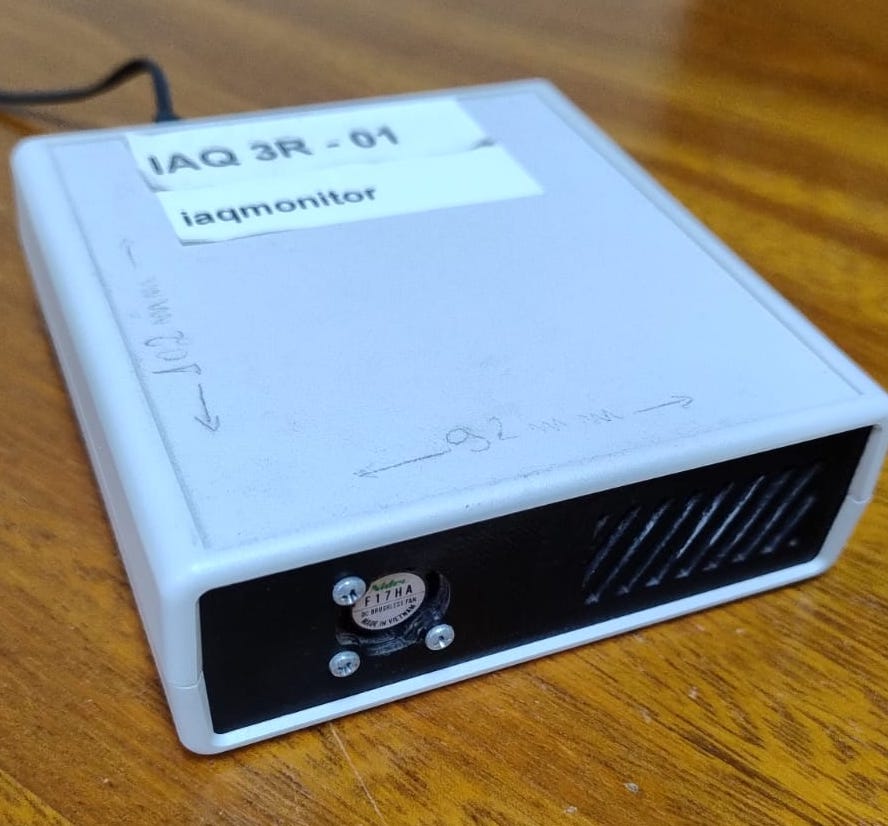}
    \caption{IAQ device used for monitoring the air quality in the rooms.}
    \label{fig:air_quality_device}
\end{figure}

The IAQ sensing device (Figure \ref{fig:air_quality_device}) has an ESP32 micro-controller with integrated Wi-Fi and Bluetooth Low Energy connectivity, plus sensors for temperature, pressure, humidity, VOC, particulates 1, 2.5, 4 and 10 microns, and a CO$_2$ sensor. The devices operate at 5V and can thus be powered by a power bank or by small photovoltaic cells. It uses WiFi to periodically transmit all probed sensor data to the \INSPIRE service, but also backups all data to a local SD card.

\subsection{Photovoltaic BLE Beacons}

We developed custom photovoltaic-powered BLE beacons to enable the indoor localization infrastructure.
The primary design objective was to create a reliable, autonomous ESP32-based unit capable of operating under a wide range of indoor lighting conditions without requiring external power sources.

The power management system consists of a photovoltaic cell connected to a CN3791 solar charge controller, which feeds a 18650 lithium-ion battery.
This configuration ensures continuous operation even in low-light environments by buffering energy in the battery, drawing inspiration from the technical foundations of energy-aware IoT gateways \cite{Talavera:SBRC:16}, which prioritize balancing data transmission with the strict energy constraints of harvested-energy environments.
To maximize autonomy, the ESP32 firmware implements an aggressive duty cycle: the device enters deep sleep (consuming approximately 5mA) for 60 seconds, followed by a 1-second wake period (peaking at 280mA) to transmit BLE advertisements.

\begin{figure}[htbp]
    \centering
    \begin{subfigure}{0.6\textwidth}
        \centering
        \includegraphics[width=\textwidth]{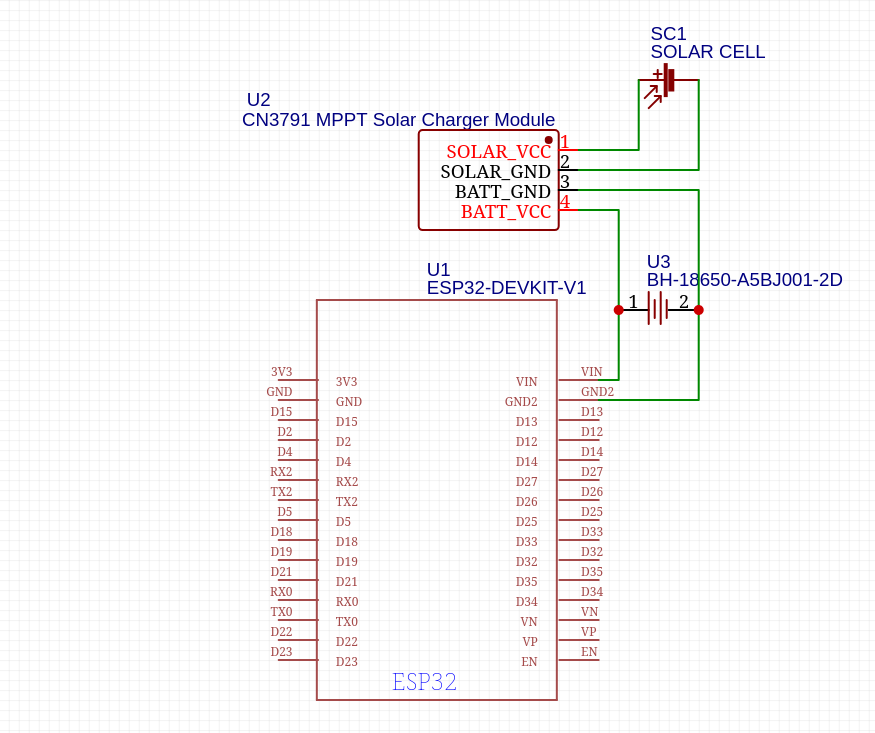}
        \caption{Circuit diagram of the BLE Beacon}
        \label{fig:beacon_diagram}
    \end{subfigure}

    \par\vspace{1em}

    \begin{subfigure}{0.85\textwidth}
        \centering
        \begin{minipage}{0.35\textwidth}
            \centering
            \includegraphics[width=\textwidth]{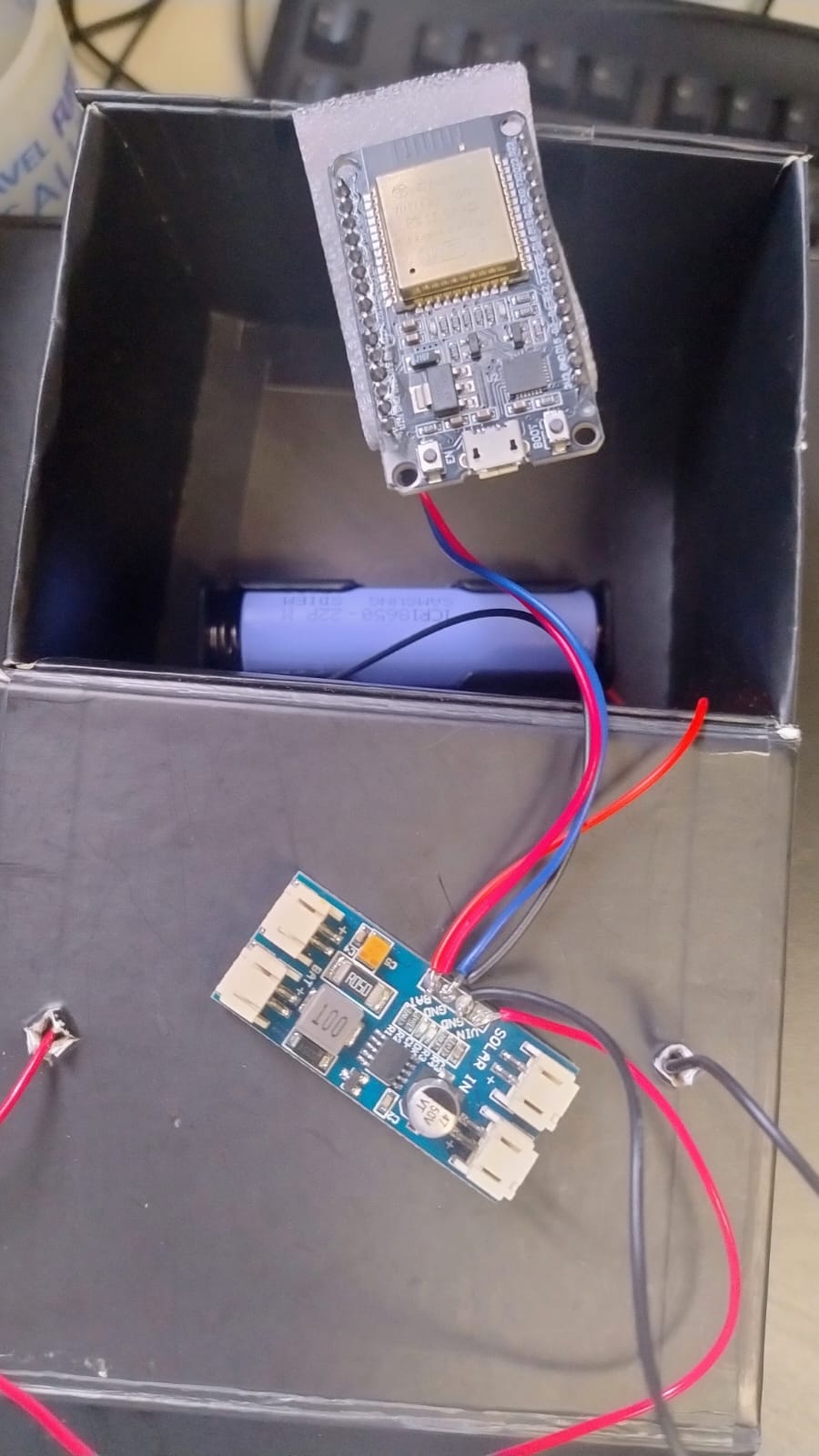}
        \end{minipage}
        \begin{minipage}{0.35\textwidth}
            \centering
            \includegraphics[width=\textwidth]{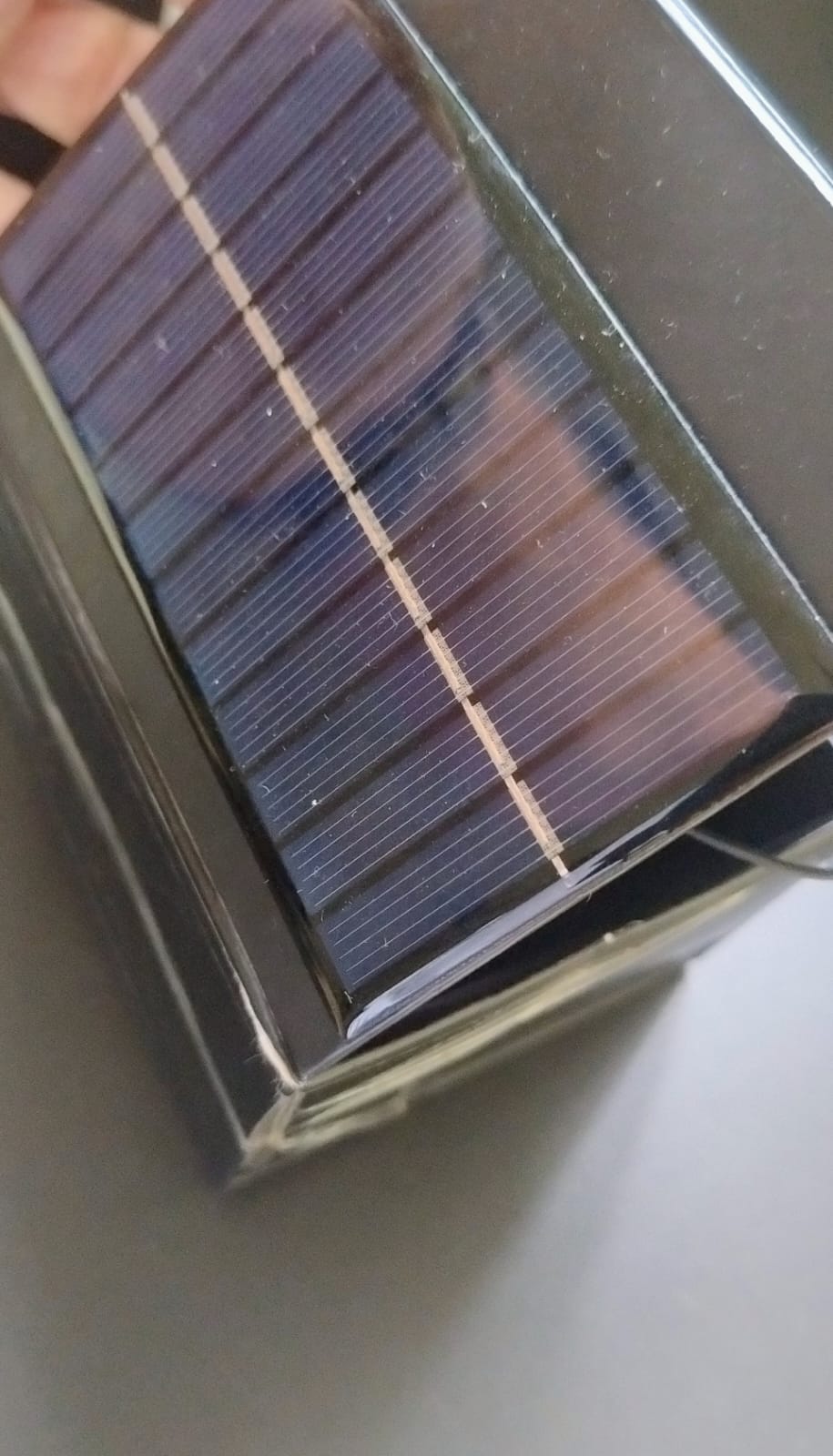}
        \end{minipage}

        \caption{Prototype of the photovoltaic BLE Beacon and solar cell}
        \label{fig:beacon_prototype}
    \end{subfigure}

    \caption{Photos and circuit diagram used in the first beacon prototype}
    \label{fig:Beacon_diagram}
\end{figure}

\subsection{EnQyMo Software}
The EnQyMo system consists of several re-engineered modules, extensions, and additional services of the original ContextNet middleware - both in the mobile edge and the backend/server/cloud components. Together, all these extensions implement the monitoring and alert notification intelligence and the focused notification of only the users directly affected by the unhealthy environmental conditions. In the remainder of this section, we discuss these enhancements and extensions in detail.

\subsubsection{Flutter-ready Mobile-Hub components}

The original \mhub\ implementation was initially developed using native Kotlin code and was aimed at Android-based devices. 
To facilitate the multi-platform deployment of the EnQyMo application across both Android and iOS environments from a single codebase, the native component was successfully encapsulated and migrated into a reusable Flutter Plugin. This architectural shift was crucial for leveraging Flutter's unified user interface capabilities while preserving the high performance and direct, low-level API access of the native \mhub\ core, as well as allowing for an easier future integration with an iOS implementation of \mhub .

The conversion process relied entirely on the Flutter Platform Channel mechanism to facilitate asynchronous communication between the Dart (Flutter) layer and the existing Kotlin code of the mhub\ components. 

For discrete function calls from Flutter to native, a standard Method Channel was established. When the Flutter application needs to control the \mhub, such as initializing its operation, it invokes a method (e.g., "startMobileHub") which is intercepted by the Kotlin onMethodCall callback. Arguments like the ipAddress and port are passed through the channel, enabling the Kotlin layer to immediately configure and initialize the native \mhub service with integrated communication protocols such as MrudpWLAN, the BLE-based BleWPAN, and the AsperCEP Complex Event Processing component \cite{Luckham:2001}. This request-response pattern ensures core commands like "stopMobileHub" and the runtime update of mobile context via "updateContext" are executed precisely in the native environment.

To handle continuous streams of incoming sensor data and \mhub\ status updates, multiple Event Channels were implemented (onMessageReceivedChannel, onBleDataReceivedChannel, and onScanningStateChangedChannel) in MobileHubPlugin. 

A key component, the onMessageReceivedHandler, demonstrates this process: it uses a reactive approach, subscribing to native MobileHubEvent. NewMessage events. As soon as a message is processed by the \mhub, its payload is immediately forwarded to the Flutter application via the EventChannel.EventSink MobileHubPlugin. This architecture guarantees that real-time sensor data, crucial for the system's function, is instantly propagated from the low-level native logic up to the high-level Dart application layer.

\subsubsection{Symbolic Location Map}
Instead of relying on geo-coordinates or noise-sensitive indoor-positioning techniques, the platform adopts a symbolic zone-based strategy to ensure predictability, resilience, and minimal calibration overhead. While other modern approaches explore decentralized location-based services and geofences on the blockchain \cite{Victor:ICDMW:2018} to manage spatial boundaries, our symbolic approach focuses on robustness and operational stability in indoor settings by operating as the abstraction layer that converts the physical BLE-beacon infrastructure into a logical zone-based presence model.

Each monitored room contains one or more BLE beacons broadcasting periodic advertisements with their unique identifiers (UUID). The continuously running M-Hub on user’s smartphones makes Bluetooth scans and locally records all the visible beacons and their respective signal strengths. This {\em beacon observation set} forms a contextual snapshot representing the perceived neighborhood of the smartphone at every moment.

The Group Definer consolidates this snapshot through deterministic symbolic-mapping rules: the detected beacon set is matched against the reference table of installed beacons, allowing the system to infer the corresponding room. This inference does not aim for absolute coordinates but rather seeks unambiguous identification of the functional zone. The output is a streamlined, actionable statement: “user present in room X.”

By operating exclusively on symbolic names of locations, the system eliminates the need for complex localization models, reduces infrastructure dependencies, and ensures operational stability. 

Through this mapping process, the Group Definer generates user groups aligned with their real-time locations, enabling the automatic generation of  targeted alerts only for workers who are effectively exposed to adverse environmental conditions.

\section{Intelligent Scientific Publication Information Retrieval Engine (\INSPIRE )}
\label{sec:INSPIRE}
\begin{figure*}[htbp]
    \centering
    \includegraphics[width=1\textwidth]{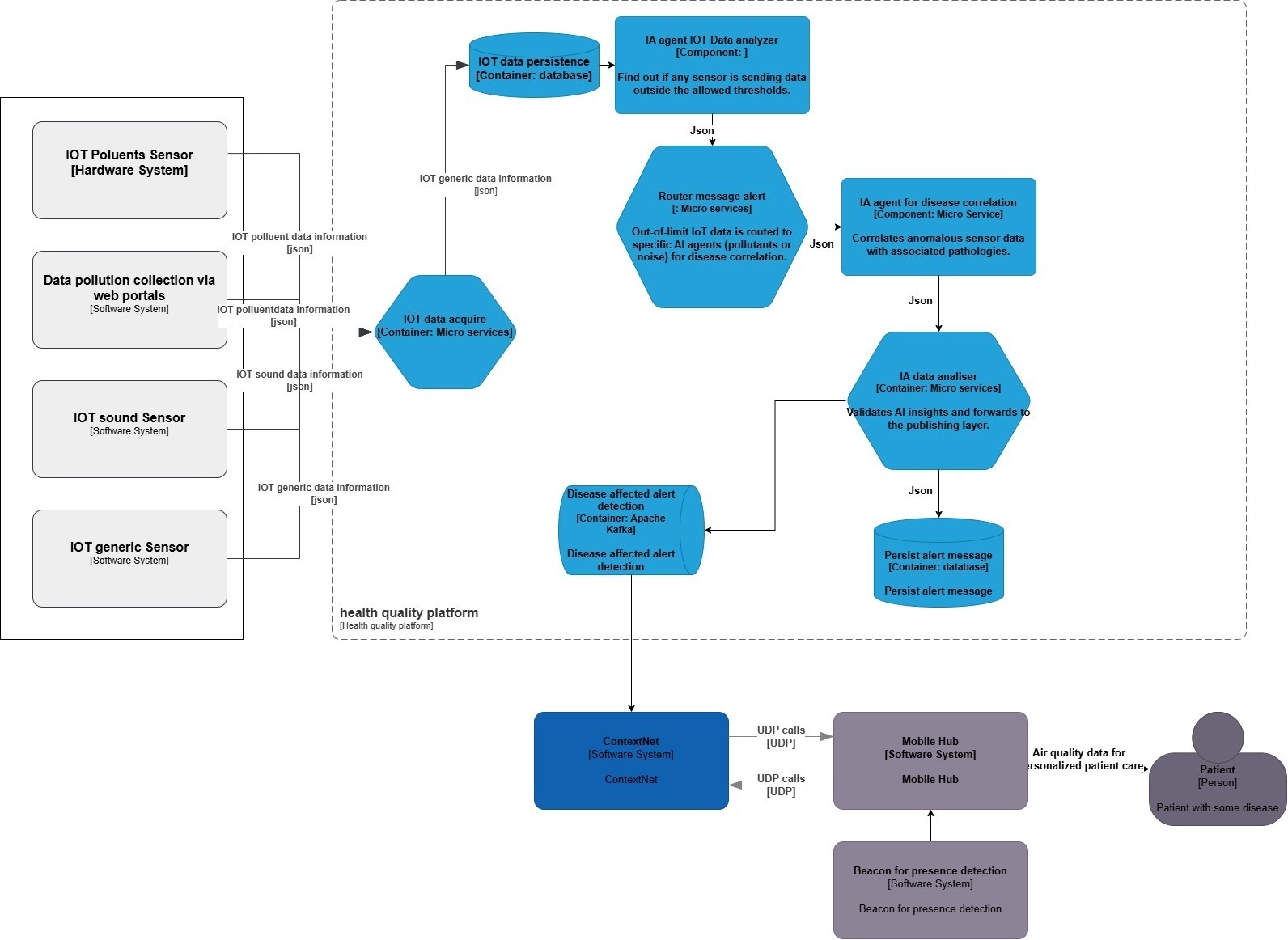}
    \caption{Architecture of \INSPIRE\ within the EnQyMo System}
    \label{fig:gailert_architecture}
\end{figure*}

At the heart of EnQyMo is an \INSPIREfull~(\INSPIRE). For each possible condition (i.e. air quality, particulates, gases, high-pitched noise and other harmful sounds, degree of vibration, etc.), the system executes  \INSPIRE\ service. To identify harmful intensities/concentration values, \INSPIRE\ utilizes an LLM trained on several scientific articles about the corresponding pollutants and their associated health problems and acute/chronic diseases. Through this learning process, a vector-based model is created that triggers corresponding alarms for users located in those unhealthy environments. This use of LLMs has several advantages: it can be easily adjusted for other environmental elements and can be constantly updated with new scientific findings, thereby improving and refining the analysis of sensory data received from environmental sensors.

The Retrieval of Information from Technical Academic Sources (\INSPIRE) is the core functional module of the Health Quality Platform.

It is composed of several specialized microservices and AI agents that work in tandem to transform raw sensor data into actionable health alerts. (Figure \ref{fig:gailert_architecture})
The components are detailed below:

\paragraph{IoT Data Acquire}
This microservice is responsible for receiving data from various sources. It normalizes the incoming data streams and persists them into a  database of past and current sensor readings, ensuring a consistent data format for downstream analysis.

\paragraph{IA Agent IoT Data Analyzer}
This agent analyzes a sample of readings performed by various sensors over a configurable time interval (e.g. 2 minutes). 
It queries the historical database for all information collected within the configured time interval and identifies if any sensor has obtained samples outside the standard acceptable by health and occupational safety regulations. If a pollutant is found to be outside the acceptable range, the agent generates an alert message for the next AI agent, indicating the sensor, the pollutant, and the criticality level. This message is sent to a microservice that routes it to the appropriate specialized agent.

\paragraph{Router Message Alert}
This component is responsible for routing the pollutant alert messages to AI agents specialized in analyzing specific data types. For example, it directs air quality issues to the Air Quality Agent and noise issues to the Sound Quality Agent.

\paragraph{IA Data Analyzer (Persistence and Publication)}
The generated analysis message is persisted in the database to create a knowledge base for quality monitoring, recording all alerts sent to ContextNet. This historical data is crucial for future root cause analysis of diseases and identifying locations or sensors that generate the most health risk alerts. After persistence, the message is published to the \texttt{AppModel} topic within ContextNet for dissemination to Mobile Hubs.

\paragraph{IA Agent for Disease Correlation}
This is an advanced AI agent capable of correlating which diseases may affect a patient or collaborator exposed to the detected pollutant. This agent utilizes a RAG-based knowledge base of scientific research documents to assist in disease correlation. These documents are previously processed into a vector database to aid the AI in accurately retrieving "Disease $\times$ Pollutant" correlations. The agent's response is formatted as a JSON message, as shown in the following listing.

\begin{figure}[htbp]
    \centering
    \includegraphics[width=1.1\textwidth]{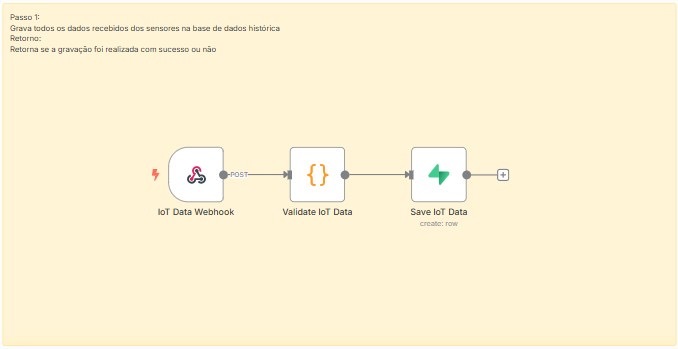}
    \caption{Architecture of \INSPIRE\ within the EnQyMo System - IOT data archive}
    \label{fig:gailert_architecture}
\end{figure}

\begin{figure}[htbp]
    \centering
    \includegraphics[width=1.1\textwidth]{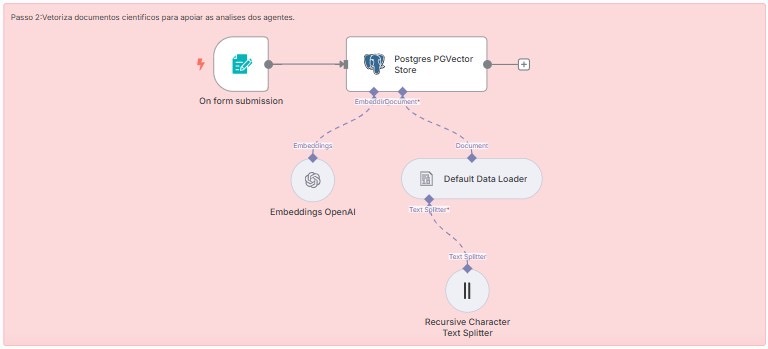}
    \caption{Architecture of \INSPIRE\ within the EnQyMo System - Vectoring knowledge Base}
    \label{fig:gailert_architecture}
\end{figure}

\begin{figure}[htbp]
    \centering
    \includegraphics[width=1.1\textwidth]{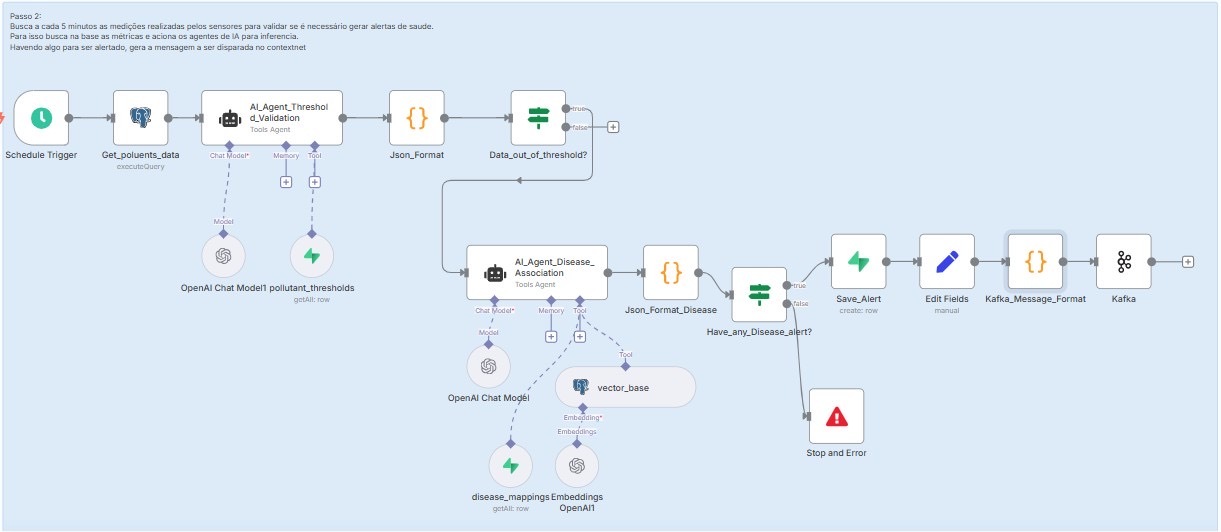}
    \caption{Architecture of \INSPIRE\ within the EnQyMo System - IA process}
    \label{fig:gailert_architecture}
\end{figure}
{\small
\begin{figure}
    \centering
    \begin{Verbatim}[fontsize=\scriptsize]
        [ 
          { 
            "analisys": { 
              "alert_id": "alert_20250813121755894", 
              "timestamp": "2025-08-13T12:17:55.894-03:00", 
              "sensores": [ 
                { 
                  "sensor_id": "IAQ_6227821", 
                  "poluentes": [ 
                    { 
                      "poluente": "pm25", 
                      "risk_level": "moderate", 
                      "affected_diseases": { 
                        "disease": [ 
                          "asthma", 
                          "bronchitis", 
                          "respiratory irritation" 
                        ] 
                      } 
                    }, 
                    { 
                      "poluente": "pm4", 
                      "risk_level": "high", 
                      "affected_diseases": { 
                        "disease": [ 
                          "respiratory irritation", 
                          "mild systemic inflammation" 
                        ] 
             ...} 
          } 
        ]
    \end{Verbatim}
    \caption{Example of the json sent as an alert}
    \label{fig:alert_json_example}
\end{figure}
}

Figure~\ref{fig:gailert_architecture} demonstrates the implementation of the agent architecture using N8N as an orchestrator. The solution is divided into 3 parts:

\begin{itemize}
\item In the yellow rectangle, we have the module responsible for acquiring data from the sensors that sent it to the published endpoint and then persisting it in a relational database.
This persisted data will be used by the main workflow, as seen in the figure, in the blue rectangle.

\item In the pink rectangle, the module is responsible for  receives academic documents, technical standards, or any material that is relevant to be used as a knowledge base for the AI agent for analysis that will be performed in the main flow by the agent for correlating alerts and diseases that may be associated.
These documents will be stored in a vector database that will be used by the agent as a RAG.

\item The blue rectangle is the main flow for disease detection and correlation.

\end{itemize}

In general, the workflow follows these steps:

\begin{enumerate}[label = \textbf{Step \arabic*}) ]

\item This flow is scheduled from time to time, retrieving the data sent and persisted by the sensors, as described above.

\item A first AI agent searches for received alerts that are above a threshold above normal. To do this, it uses a knowledge base containing the enabled technical standards on what is being measured. In this case, pollutants. 

If there is an indication of alerts above the permitted level, the agent will generate a JSON message indicating the sensor that generated the alert, which pollutant is above the permitted level, and what level above the permitted level it is at.
\\
\item If there are measurements that are above the permitted level, the second AI agent is activated with the responsibility of correlating the pollutant that is above the permitted level with diseases that this exposure can cause to a person. 

For this, it has two tools that will be used as a knowledge base. These are the vector documents included by the process described above and a database with information collected by research and knowledge of experts on the subject.

The agent makes these correlations in two steps, the first being to search for the most assertive data from experts and technical standards. The second analysis, in order to verify and refine the information obtained in the first step, uses the vector documents.
After these analyses, the agent will generate a message in JSON format (shown in Figure~\ref{fig:alert_json_example}) containing the sensor, the altered pollutant, and the possible associated diseases.
\\
\item This message generated by the AI agent is then formatted and sent to the contextNet in order to be directed to patients who are in the sensor's location and who are at risk of the identified diseases.

\end{enumerate}

\section{Early Experiments}
\label{sec:xperiments}
To validate the feasibility and end-to-end functionality of the EnQyMo system, we conducted a proof-of-concept experiment in the lab NITAS (Núcleo de Inovação em Tecnologias e Aplicações para a Saúde) at PUC-Rio. The experimental setup involved a complete deployment of the hardware and software pipeline described in Section \ref{sec:system}, specifically designed to demonstrate environmental monitoring, risk inference, and targeted alert delivery.

\subsection{Environmental Data Acquisition}
The environment was monitored using the IAQ 3R-01 device, positioned on a desk of laboratory to capture air quality data. During the experiment, the device successfully streamed real-time sensor data to the \INSPIRE\ , which was visualized on a dashboard displaying critical metrics. As observed in our tests, the sensors captured Carbon Dioxide (CO$_{2}$) levels at 998 ppm, temperature at 24.3$^{\circ}$C, relative humidity at 66.1\%, and concentrations of particulate matter (PM1.0, PM2.5, PM10).

\subsection{Workflow Automation and Health Inference}
The data processing pipeline was orchestrated using N8N workflows. We simulated a scenario where pollutant levels exceeded safety thresholds. The experiment demonstrated the system's ability to:
\begin{enumerate}
    \item Receive sensor data at the node executing the \INSPIRE ;
    \item Analyze the data against configured safety thresholds;
    \item Trigger the AI analysis agent to identify associated diseases (e.g., asthma, bronchitis, respiratory irritation);
    \item Publish a structured alert message to the Kafka queue.
\end{enumerate}

\subsection{Targeted Alert Delivery}
To verify the \textit{Groupcast} and presence detection capabilities, three  smartphones (Android devices) running the Mobile-Hub application were placed in the same room. The devices detected the Bluetooth Low Energy (BLE) beacons, successfully identifying their presence within the lab room.

\begin{figure}[htbp] 
    \centering
    \begin{subfigure}[b]{0.5\textwidth}
        \centering
        \includegraphics[width=\textwidth]{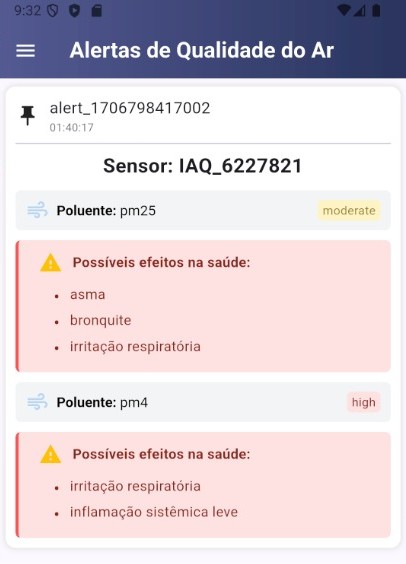} 
        \caption{Smartphone receiving alert}
        \label{fig:alert_screen}
    \end{subfigure}
    \hspace{0.1\textwidth}
    \begin{subfigure}[b]{0.5\textwidth}
        \centering
        \includegraphics[width=\textwidth]{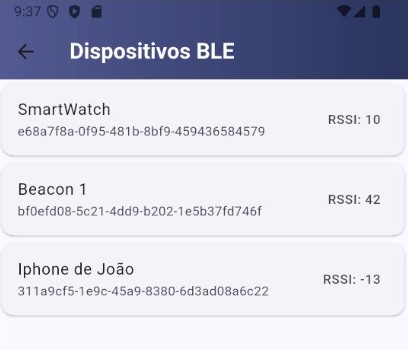}
        \caption{Smartphone detecting BLE beacons}
        \label{fig:beacon_screen}
    \end{subfigure}
    
    \caption{Screenshots of the Mobile Hub application executing the alert and detection logic.}
    \label{fig:MHUB_flutter}
\end{figure}

Upon processing the high-pollution event, the system broadcasted a targeted message. All three devices simultaneously received the JSON-formatted alert notification containing the pollutant details (PM2.5), the risk level ("moderate"), and the inferred health risks. This confirmed the correct operation of the ContextNet GroupDefiner and the Mobile-Hub's ability to act as an edge gateway for alerting users in hazardous environments. \cite{Talavera:2015}

\subsection{Solar Beacon Autonomy}
Finally, we validated the hardware design of the photovoltaic BLE beacons. The prototype, utilizing a CN3791 solar charge controller, an ESP32 microcontroller, and an 18650 battery, successfully transmitted advertising packets used for RSSI-based proximity detection. The varying RSSI values were displayed on the mobile application during the test, confirming that the system can distinguish distance and proximity to specific zones even when powered by harvested energy.

\subsection{Experimental Limitations}

While the results described above confirm the functional viability of the EnQyMo architecture, we acknowledge the limitations inherent in this early proof-of-concept. The experiments were conducted in a controlled laboratory setting using simulated pollution events to trigger the alert logic, rather than during longitudinal deployment in a complex, densely occupied building. Consequently, this phase did not fully evaluate long-term challenges such as sensor drift, the impact of physical obstructions on BLE signal stability in varying layouts, or the varying efficiency of the photovoltaic beacons under suboptimal lighting conditions over extended periods. Future validation phases will move beyond the laboratory to a multi-room pilot study, focusing on the system's scalability, the durability of the energy-harvesting hardware, and the practical utility of the notifications in a real-world workflow.

\section{Simulation based experiments}
\label{sec:simulation}
\subsection{Simulation Environment and Hardware Setup}
To accurately assess the scalability and responsiveness of the EnQyMo platform under concurrent load, we conducted a series of stress tests using simulated mobile nodes rather than physical smartphone devices. The simulations were executed on a Lenovo Legion 5i Gen 10 equipped with 32 GB of RAM and an NVIDIA RTX 5070 GPU with 8 GB of VRAM and an Intel Core Ultra 7 255HX. 

Employing Java-based node emulations, rather than the actual physical mobile devices utilized in early experiments, allowed us to isolate the core performance of the ContextNet gateway and the alert distribution mechanism. This approach removes the unpredictable network anomalies, battery throttling, and wireless interference inherent to physical testing environments, providing a clean baseline for evaluating the computational scalability of the platform.

\subsection{Stress Test Methodology and Node Mobility}

The stress test 
was driven by a custom Java application (\texttt{StressTestRunner}) designed to launch $N$ autonomous mobile nodes against the ContextNet gateway. To ensure accurate simulation of independent clients and to avoid thundering-herd issues on the gateway, the runner dynamically generated a unique UUID for each node, initializing them with a 100-millisecond stagger. 

While the simulated nodes are capable of operating in a \textit{dynamic mode}, autonomously hopping between different Bluetooth Low Energy (BLE) beacons to simulate the mobility of the workforce, the metrics for this evaluation were obtained exclusively using the \textit{static mode}. In static mode, all simulated nodes were pinned to a specific beacon representing a single physical zone. 

This methodological choice is crucial for accurate measurement. Node mobility does not significantly change the computational cost of context monitoring (group matching) at the gateway, since the context updates regardless of wether the node has moved or not. 
However, dynamic hopping introduces fluctuating group sizes, which leads to an unpredictable and constantly varying number of expected Acknowledgments (ACKs) for any given alert.
By pinning the nodes statically, we eliminated this uncontrolled variable, guaranteeing exactly $N$ nodes per targeted group to reliably measure the precise distribution latency and system overhead.

\begin{figure}
    \centering
    \includegraphics[width=0.9\linewidth]{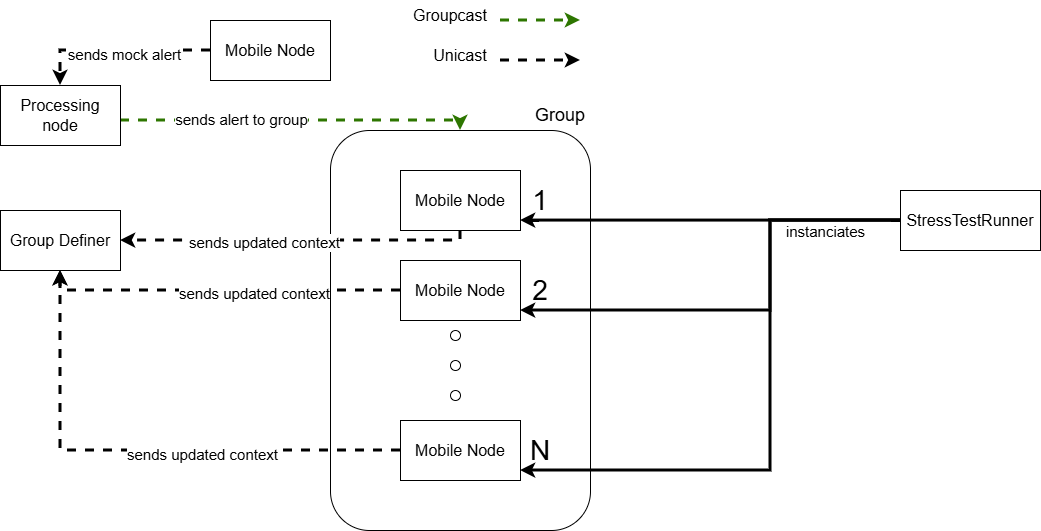}
    \caption{Architecture of the simulation environment}
    \label{fig:stress_test_diagram}
\end{figure}

Figure \ref{fig:stress_test_diagram} illustrates the simulation architecture. The simulated mobile nodes continuously send each of their updated contexts to the Group Definer, while another separate simulated mobile node, not in the same group, sends a mocked alert to the Processing Node, simulating the alert from INSPIRE.

\subsection{Latency Analysis and Results}
The primary metric evaluated during the stress tests was the Average Round Trip Time (RTT), measured in milliseconds (ms). The RTT encompasses the time required to dispatch localized alerts from the gateway and receive ACKs back from the target nodes. The stress test recorded alert distribution events scaling from 5 to 100 concurrent target nodes within a single zone.

\begin{figure}[htbp]
  \centering
  \includegraphics[width=0.8\textwidth]{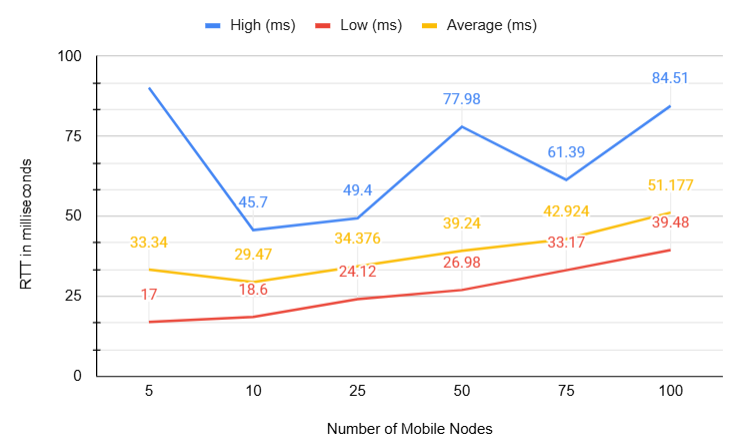}
  \caption{Time series graph illustrating the Low, High and Average Round Trip Time (RTT) in milliseconds as the number of concurrent target nodes increases}
  \label{fig:series_graph}
\end{figure}

An analysis of the snapshot data reveals that the system scales efficiently while maintaining extremely low latency. As summarized in Table \ref{tab:rtt_results}, the system consistently achieved average RTTs well below the 100 ms threshold, even at peak capacities.

\begin{table}[htbp]
\centering
\caption{Summary of Alert Distribution Round Trip Times (RTT) in Static Mode}
\label{tab:rtt_results}
\begin{tabular}{cccc}
\toprule
\textbf{Target Nodes (ACKs)} & \textbf{Min RTT (ms)} & \textbf{Max RTT (ms)} & \textbf{Mean RTT (ms)} \\
\midrule
5   & 17.00 & 90.20 & 33.34 \\
10  & 18.60 & 45.70 & 29.47 \\
25  & 24.12 & 49.40 & 34.38 \\
50  & 26.98 & 77.98 & 39.24 \\
75  & 33.17 & 61.39 & 42.92 \\
100 & 39.48 & 84.51 & 51.18 \\
\bottomrule
\end{tabular}
\end{table}

\begin{itemize}
    \item \textbf{Low-Density Alerts (5 to 10 nodes):} The platform demonstrated near-instantaneous dissemination, with mean RTTs hovering around 29 to 33 ms. A single outlier at 90.20 ms for a 5-node distribution was observed, likely due to initial JVM warmup overhead.
    \item \textbf{Medium-Density Alerts (25 to 50 nodes):} As the number of target nodes increased, the average RTT remained highly stable, increasing only marginally to a mean of 39.24 ms for 50 simultaneous ACKs.
    \item \textbf{High-Density Alerts (75 to 100 nodes):} At peak tested capacity, disseminating alerts to 100 distinct nodes yielded a mean RTT of 51.18 ms, with the absolute maximum delay recorded at 84.51 ms. 
\end{itemize}

\begin{table}[htbp]
\centering
\caption{Distribution of RTTs (ms) for Box Plot Analysis}
\label{tab:boxplot_data}
\begin{tabular}{ccccc}
\toprule
\textbf{Target Nodes} & \textbf{Min} & \textbf{Median} & \textbf{p95} & \textbf{Max} \\
\midrule
5   & 17.00 & 24.00 & 69.77 & 90.20 \\
10  & 18.60 & 25.70 & 43.77 & 45.70 \\
25  & 24.12 & 29.74 & 49.24 & 49.40 \\
50  & 26.98 & 33.82 & 62.06 & 77.98 \\
75  & 33.17 & 39.95 & 55.13 & 61.39 \\
100 & 39.48 & 46.44 & 73.67 & 84.51 \\
\bottomrule
\end{tabular}
\end{table}

\begin{figure}[htbp]
  \centering
  \includegraphics[width=1\textwidth]{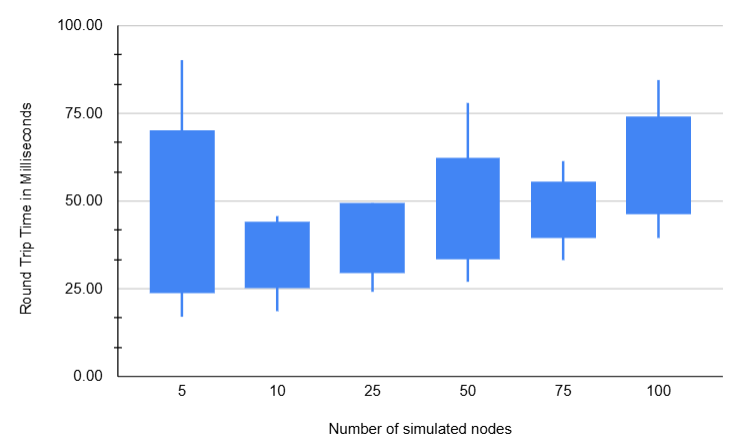}
  \caption{Box plot showing the variance and distribution of RTTs across different target node cluster sizes.}
  \label{fig:boxplot}
\end{figure}

The box plot analysis (Table \ref{tab:boxplot_data} and Figure \ref{fig:boxplot}) provides deeper insight into the distribution and variance of the RTTs. The median response times exhibit a stable, modest increase as the node count scales, rising linearly from 24.00 ms (5 nodes) to 46.44 ms (100 nodes). Furthermore, the 95th percentile (p95) values remain consistently low, demonstrating that the vast majority of alert distributions complete predictably without heavy fluctuations. For example, even at peak tested capacity (100 nodes), 95\% of the distribution events complete in under 73.67 ms. While there are occasional maximum spikes, such as the aforementioned 90.20 ms outlier at 5 nodes and 84.51 ms at peak load, this tight clustering underscores the reliability of the gateway's asynchronous processing and the event-driven architecture of the ContextNet middleware.

Overall, the EnQyMo platform demonstrated robust scalability. The results confirm that the underlying architecture is highly capable of supporting near real-time, localized alert generation for large populations within monitored indoor spaces without suffering from network or processing bottlenecks.

\section{Conclusion}
\label{sec:conclusion}
The EnQyMo project successfully demonstrated the integration of low-cost IoT hardware, scalable middleware, and Generative AI to address occupational health challenges. By leveraging the ContextNet architecture and the Mobile-Hub concept, we established a robust pipeline that not only monitors environmental variables but also contextualizes them into actionable health insights.

The key contribution of this work lies in the development of a general framework that combines AI agency, LLMs and a scalable IoT middleware to shift from passive monitoring to proactive, occupant-aware alerting for occupational health. Unlike traditional systems that merely log environmental data in a cloud a and display statistics on a dashboard, EnQyMo uses agentic AI to interpret environmental sensor readings taking into account knowledge about health parameters and risks that a LLM obtains from scientific and medical literature. This allows the system to deliver localized warning alerts about specific risks—such as respiratory irritation or asthma triggers—directly to the smartphones of employees exposed to those conditions.

Furthermore, the implementation of zone-based localization using solar-powered BLE beacons ensures that alerts are spatially accurate without requiring expensive maintenance of the localization infrastructure (i.e., battery replacement) or specific and  costly indoor positioning technology, but using only BLE beacons and user smartphones. The successful migration of the Mobile-Hub components to a Flutter plugin also ensures that the solution is ready for cross-platform deployment, enhancing its accessibility for diverse workforce environments.

\subsection{Future Work}
\label{sec:future}
While the current implementation validates the core architecture, several avenues for future development have been identified:
\begin{itemize}
    \item \textbf{Expanded Environmental Metrics:} The current system focuses on air quality ($CO_{2}$ and Particulate Matter). Future iterations will integrate noise level monitoring and vibration sensors into the risk analysis pipeline, as these are also critical factors in occupational health.
    
    \item \textbf{M-Hub-Mediated Data Acquisition:} Currently, sensor data is transmitted directly to the backend via WiFi. Future work will leverage the Mobile-Hub's capability to act as a mobile gateway, as originally designed in the ContextNet architecture. In this model, smartphones will discover local sensors via WPAN (e.g., BLE) and aggregate their data before transmission. This enhances mobility and eliminates the dependence on fixed WiFi infrastructure for the sensor nodes.
    
    \item \textbf{Edge-Based Complex Event Processing (CEP):} Although the Mobile-Hub implementation currently includes the AsperCEP engine, it is not yet utilized for logic processing. Future developments will activate these CEP capabilities to process sensor streams locally on the smartphone. This will allow for "Edge AI" filtering, identifying critical anomalies or rapid pollutant spikes in real-time, thereby reducing cloud bandwidth usage and improving system responsiveness.

    \item \textbf{Long-term Autonomy Tests:} While the solar beacons function correctly in short-term tests, long-term field studies are necessary to validate the energy harvesting efficiency of the CN3791 and solar panel configuration under varying indoor lighting conditions.
\end{itemize}

\section{Acknowledgments}
\label{sec:acknowledgments}
 Language editing and grammatical corrections were assisted by Google Gemini. The authors reviewed all changes and retain full accountability for the final text.

\begin{backmatter}
\bibliographystyle{vancouver.bst}
\bibliography{main}
\end{backmatter}

\end{document}